\documentclass[%
reprint,
amsmath,amssymb,
aps,
twocolumn,
]{revtex4-2}

\usepackage{color}

\usepackage{amssymb}
\usepackage{braket}
\usepackage{graphicx}
\usepackage{dcolumn}
\usepackage{bm}

\usepackage{algorithm}
\usepackage{algpseudocode}

\usepackage{tikz}

\tikzset{
  panel label/.style={
    anchor=north west,
    font=\footnotesize,
    fill=white,
    inner sep=1.5pt,
    rounded corners=1pt
  }
}

\begin{document}

\title{Scalable 
Quantum Monte Carlo Method for 
Polariton Chemistry
via Mixed Block Sparsity and Tensor Hypercontraction Method
}

\author{Yu Zhang}
\email{zhy@lanl.gov}
\affiliation{%
Theoretical Division, Los Alamos National Laboratory, Los Alamos, NM, 87545. 
}%

\date{\today}

\begin{abstract}
We present a reduced-scaling auxiliary-field quantum Monte Carlo (AFQMC) framework designed for large molecular systems and ensembles, with or without coupling to optical cavities. Our approach leverages the natural block sparsity of Cholesky decomposition (CD) of electron repulsion integrals in molecular ensembles and employs tensor hypercontraction (THC) to efficiently compress low-rank Cholesky blocks. By representing the Cholesky vectors in a mixed format, keeping high-rank blocks in block-sparse form and compressing low-rank blocks with THC, we reduce the scaling of exchange-energy evaluation from quartic to robust cubic in the number of molecular orbitals $N$, while lowering memory from cubic toward quadratic. Benchmark analyses on one-, two-, and three-dimensional molecular ensembles (up to $\sim$1,200 orbitals) show that: a) the number of nonzeros in Cholesky tensors grows linearly with system size across dimensions; b) the average numerical rank increases sublinearly and does not saturate at these sizes; and (c) rank heterogeneity—some blocks nearly full rank and many low rank, naturally motivating the proposed mixed block sparsity and THC scheme for efficient calculation of exchange energy. We demonstrate that the mixed scheme yields cubic wall-time scaling with favorable prefactors and preserves AFQMC accuracy.
\end{abstract}

\maketitle

\section{Introduction}
\label{sec:introduction}

The interaction of molecular systems with quantized cavity photons gives rise to hybrid light-matter states, or polaritons. Experiments and theories have shown that such states can reshape chemical landscapes~\cite{Nagarajan2021JACS, GarciaVidal2021S, Hutchison2012ACIE, Schwartz2011PRL, Ebbesen2016ACR, Thomas2016ACIE, Thomas2019S,hirai_molecularPOL_ChemRev2023, Galego:2015aa, C8SC01043A, Zhang2019JCP}, alter material properties~\cite{Thomas2021, Berghuis2020JPCC, Berghuis2022AP,xu_ultrafast_NatCommun2023, Deng2010RMP}, and mediate new transport mechanisms~\cite{rozenman_PolTransport_ACSPhot2018,Allen_SWCNTpol_JPCC2022,Lindoy2022JPCL}. These discoveries have stimulated broad theoretical efforts to extend electronic structure methods into the domain of cavity quantum electrodynamics (QED)~\cite{Mandal:2023cr,ruggenthaler_QED_ChemRev2023,weight_QEDReview_PCCP2023}.
Several quantum chemistry approaches have been adapted to the Pauli-Fierz Hamiltonian~\cite{Bauman:2025ti}, including QED extensions of Hartree-Fock~\cite{haugland2020PRX, Schnappinger:2023tx, Cui:2024ul, riso2022NatComm, li_vQED_arXiv2023, Mazin:2024aa}, density functional theory~\cite{Flick2020JCP, yang2021JCP, yang_polGRADS_JCP2022, liebenthal2022JCP, liebenthal_mean-field_Arxiv2023, Vu_enhanced_JPCA2022, yang2021JCP, yang_polGRADS_JCP2022}, coupled-cluster theory~\cite{haugland2020PRX, Haugland2021JCP, Pavosevic2022JACS, deprince2021JCP, White:2020jcp}, configuration interaction~\cite{Foley:2023aa, Vu:2024aa}, QED complete active space self-consistent field (QED-CASSCF)~\cite{vu2025cavity, alessandro_complete_2025}, and quantum monte carlo methods~\cite{Weight:2024pra, Tang:2025ul, Weber:2025aa, Zhang:2025qmc}.
While these methods have enabled the first \emph{ab initio} studies of polaritonic chemistry, they face severe limitations: density functionals require new forms to capture electron-photon correlation, and coupled-cluster approaches become intractable when large photon numbers or many molecules are involved. This has restricted predictive polaritonic simulations to small systems and weak-to-moderate coupling.

Auxiliary-field quantum Monte Carlo (AFQMC) offers a systematically improvable alternative.
AFQMC has been established as an efficient many-body method for correlated electrons~\cite{Shi:2021wv}, and its generalization to electron-boson Hamiltonians provides a natural route to study polaritonic ground states with minimal uncontrolled approximations~\cite{Tang:2025ul, Zhang:2025qmc, Weber:2025aa}.
The main bottleneck lies in handling the two-electron integrals and exchange energy contributions, which scale as $\mathcal{O}(N^4)$ with the number of orbitals $N$, making direct AFQMC propagation impractical for large molecular ensembles.

In this work, we overcome this bottleneck by introducing a mixed block-sparsity and low rank factorization strategy for AFQMC.
The Cholesky-decomposed two-electron integrals exhibit block sparsity (BS) in molecular ensembles, as inter-molecular integrals vanish in the absence of direct Coulomb coupling. 
Within each block, most Cholesky tensors are further shown to be low rank and can be efficiently factorized using tensor hypercontraction (THC).
We therefore represent the two-electron integrals as a combination of block-sparse and THC-compressed tensors.
This mixed decomposition reduces the computational scaling of exchange energy evaluation to $\mathcal{O}(N^3)$ while retaining systematically controllable accuracy.
The resulting cubic-scaling AFQMC algorithm is especially advantageous in the \emph{collective coupling regime}, where many molecules coherently interact with one or a few cavity modes. In this regime, cavity-mediated dipole-dipole interactions dominate, while intra-molecular contributions remain sparse and low rank, making the mixed BS-THC representation maximally efficient.
We demonstrate that the method achieves near-full configuration interaction accuracy for small polaritonic systems and extends AFQMC simulations to molecular ensembles of practical experimental relevance.
This establishes AFQMC as a powerful and scalable tool for predictive modeling of cavity-modified chemistry and strongly correlated polaritonic matter.

The paper is organized as follows. Section~\ref{sec:theory} introduces the QED-AFQMC framework and outlines several strategies for estimating the exchange energy. Section~\ref{sec:mixed} describes our mixed scheme, which exploits both the block-sparse and low rank structures of the Cholesky tensors to achieve more efficient exchange energy estimation, along with a corresponding complexity analysis. Numerical examples and discussions are presented in Section~\ref{sec:results}. Finally, Section~\ref{sec:summary} concludes the paper.

\section{Methodology}
\label{sec:theory}

\subsection{Brief introduction to AFQMC for coupled systems}

The molecular Hamiltonian can be extended to include electron-photon interactions via the Pauli-Fierz Hamiltonian,
\begin{align}
    \hat{H}_\mathrm{PF} &= \hat{H}_\mathrm{e} + \hat{H}_\mathrm{ph} + \hat{H}_\mathrm{e-ph} + \hat{H}_\mathrm{DSE}.
\end{align}
where $\hat{H}_\mathrm{e}$ is the bare electronic Hamiltonian; the photonic Hamiltonian $\hat{H}_\mathrm{ph}$, the bilinear coupling term $\hat{H}_\mathrm{e-ph}$, and the dipole self-energy (DSE) $\hat{H}_\mathrm{DSE}$ are
\begin{align}
    \hat{H}_\mathrm{ph} &= \sum_{\alpha} \omega_{\alpha} (\hat{a}^\dagger_{\alpha}\hat{a}_{\alpha} + \frac{1}{2}), \\
    \hat{H}_\mathrm{e-ph} &= \sum_{\alpha}\sqrt{\frac{\omega_{\alpha}}{2}}(\boldsymbol{\lambda}_{\alpha} \cdot \boldsymbol{\hat{D}})(\hat{a}^\dagger_{\alpha} + \hat{a}_\alpha), \\
    \hat{H}_\mathrm{DSE} &= \sum_{\alpha} \frac{1}{2} (\boldsymbol{\lambda}_{\alpha} \cdot \boldsymbol{\hat{D}})^2.\nonumber
\end{align}
where $\boldsymbol{\lambda}_{\alpha} = \sqrt{\frac{1}{\epsilon \mathcal{V}}} \mathbf{\hat{e}}_\alpha = \lambda_\alpha \mathbf{\hat{e}}_\alpha$ is the electron-photon coupling strength of the $\alpha^\mathrm{th}$ photon mode with unit polarization direction $\mathbf{\hat{e}}_\alpha$, $\epsilon$ is the dielectric constant of the medium, and $\mathcal{V}$ is the confining mode volume. The molecular dipole operator includes both electronic and nuclear components, $\hat{\boldsymbol{D}} = -\sum_i^{N_\mathrm{e}}e\mathbf{\hat{r}}_i + \sum_I^{N_\mathrm{N}} Z_I \mathbf{R}_I$,
where $Z_I$ is the charge of the $I^\mathrm{th}$ nucleus.
As shown in previous work~\cite{Zhang:2025qmc}, the DSE Hamiltonian depends solely on electronic operators, leading to modifications of the one-electron integrals (OEI) and electron-repulsion integrals (ERI). Consequently, the final PF Hamiltonian can be written as
\begin{align}\label{eq:finalpf}
    \hat{H}_\mathrm{PF} =& \sum_{pq}^{N} \tilde{h}_{pq}\hat{c}_p^\dag \hat{c}_q + \frac{1}{2}\sum_{pqrs}^{N} \tilde{V}_{pqrs} \hat{c}_p^\dag \hat{c}_q^\dag \hat{c}_r \hat{c}_s \\
    &+ \sum_\alpha^{P} \omega_\alpha (\hat{a}^\dag_\alpha\hat{a}_\alpha + \frac{1}{2})
    \nonumber\\ 
    &+ \sum_{pq}^{N}\sum_\alpha^{P}\sqrt{\frac{\omega_\alpha}{2}} g_{pq}^\alpha (\hat{a}^\dagger_\alpha + \hat{a}_\alpha)\hat{c}_p^\dag \hat{c}_q\nonumber
\end{align}
where $\tilde{h}_{pq} = h_{pq} - \frac{1}{2}\sum_\alpha^{P}g_{pr}^\alpha g^\alpha_{rq}$ and $\tilde{V}_{pqrs} = V_{pqrs} + \sum_\alpha^{P} g_{pq}^\alpha g_{rs}^\alpha$ are the DSE-modified OEI and ERI. 
The final Pauli-Fierz (PF) Hamiltonian in dipole gauge consists of the bare electronic and photonic Hamiltonians and a bilinear light-matter coupling term. $N$ and $P$ are the number of orbitals and photonic modes, respectively.  

\subsubsection{Monte Carlo (MC) Hamiltonian}
The AFQMC formalism requires rewriting the original Hamiltonian into the format of a so-called Monte Carlo Hamiltonian,
\begin{equation}
    \hat{H}_{mc} = \hat{T} + \frac{1}{2}\sum^{N_\gamma}_\gamma \hat{L}^2_\gamma + C,
\end{equation}
which consists of a one-body operator $\hat{T}$, squares of one-body terms $\hat{L}_\gamma$, and a constant shift. 
With the Cholesky decomposition (CD)~\cite{Aquilante:2007uj, Beebe:1977ve, Koch:2003vj, Roeggen:2008wh}, the ERI is rewritten as
\begin{equation}
    V_{pqrs} = \sum^{N_\gamma}_\gamma L^e_{\gamma, pq} L^{e,*}_{\gamma, rs}.
    \label{eq:cd}
\end{equation}
where $\{L^\gamma\}$ are rank-three Cholesky tensors. Here, $N_\gamma$ is the number of Cholesky tensors needed to reach a given accuracy, typically scaling linearly with $N$, which can be controlled by the threshold ($\epsilon$) of the low rank approximation. 
On the other hand, the bilinear coupling term can be rewritten into the MC Hamiltonian via the decomposition
\begin{equation}
  \hat{F}_\alpha \hat{Q}_\alpha = \frac{(\hat{F}_\alpha + \hat{Q}_\alpha)^2 - (\hat{F}_\alpha - \hat{Q}_\alpha)^2}{4}.
\end{equation}
where $\hat{F}_\alpha \equiv \sum_{pq} \sqrt{\frac{\omega_\alpha}{2}}g^\alpha_{pq}\hat{c}^\dagger_p \hat{c}_q$ and $\hat{Q}_\alpha= (\hat{a}^\dagger_\alpha+\hat{a}_\alpha)$.
After incorporating the electron-boson interaction Hamiltonian and reorganizing the one-body and two-body interactions, the final MC format of the original PF Hamiltonian reads~\cite{Zhang:2025qmc}
\begin{equation}\label{eq:mcHam}
  \hat{H}_{MC}
  = \hat{T} + \frac{1}{2}\sum^{N_\gamma}_{\gamma} \hat{L}^{e,2}_\gamma
  + \frac{1}{2}\sum^{2P}_{\gamma'} \hat{L}^{ep,2}_{\gamma'} + \hat{H}_{ph}.
\end{equation}
Where $\hat{L}^{ep}_{\gamma'} \in \left\{ \frac{\hat{F}_\alpha + \hat{Q}_\alpha}{\sqrt{2}}, \frac{i(\hat{F}_\alpha - \hat{Q}_\alpha)}{\sqrt{2}} \right\}$ are the operators resulting from the decomposition of the bilinear coupling term.
The effective kinetic operator is $\hat{T}= \sum_{pq} T_{pq}\hat{c}^\dag_p \hat{c}_q$ and
\begin{equation}
    T_{pq} = \tilde{h}_{pq}
    - \frac{1}{2}\sum_{\gamma r} L^{e,*}_{\gamma,pr}L^e_{\gamma,rq}.
\end{equation}

\subsection{AFQMC Scheme for Molecular Quantum Electrodynamics}

AFQMC computes the ground state via imaginary-time evolution:
\begin{equation}
    \ket{\Psi_0} \propto \lim_{\tau \rightarrow \infty} e^{-\tau \hat{H}} \ket{\Psi_T}.
\end{equation}
The ground state $\ket{\Psi_0}$ of a many-body Hamiltonian \(\hat{H}\) can be projected from any trial wavefunction \(\ket{\Psi_T}\) that satisfies the non-orthogonality condition \(\braket{\Psi_T | \Psi_0} \neq 0\). 
In practice, the imaginary-time evolution is discretized into a sequence of small time steps $\ket{\Psi^{(n+1)}} = e^{-\Delta\tau \hat{H}} \ket{\Psi^{(n)}}$.
When $\Delta\tau$ is sufficiently small, the one-body and two-body terms in the evolution operator can be further factorized by the Suzuki-Trotter decomposition~\cite{Suzuki:1976aa}. For instance, the widely used symmetric decomposition is
\begin{align}
    e^{-\Delta\tau \hat{H}_{\text{MC}}} &\approx e^{-\frac{\Delta\tau}{2} \hat{T}}
    e^{-\frac{\Delta\tau}{2} \hat{H}_{\text{ph}}}
    e^{-\Delta\tau \sum_\gamma \frac{\hat{\mathcal{L}}^2_\gamma}{2}}
    \nonumber \\
    & \quad \times e^{-\frac{\Delta\tau}{2} \hat{H}_{\text{ph}}}
    e^{-\frac{\Delta\tau}{2} \hat{T}} e^{-\Delta\tau C} + \mathcal{O}(\Delta \tau^3).
\end{align}
The action of one-body propagators ($e^{-\frac{\Delta\tau}{2} \hat{T}}$ and $e^{-\frac{\Delta\tau}{2} \hat{H}_{\text{ph}}}$) on the walker wavefunction is straightforward and numerically efficient. The primary computational challenge arises from the two-body propagator.

After rewriting the Hamiltonian in the MC format, applying the Hubbard-Stratonovich (HS) transformation~\cite{Hubbard:1959vo} to the quadratic term in the imaginary time evolution operator ($\exp[-\hat{L}^2_\gamma]$) converts them into stochastic couplings to one-body propagators, 
\begin{equation}
    e^{-\Delta\tau \sum_\gamma \hat{\mathcal{L}}^2_\gamma / 2}
    = \prod_\gamma \int dx_\gamma \frac{1}{\sqrt{2\pi}}
    e^{-x^2_\gamma / 2} e^{x_\gamma \sqrt{-\Delta\tau} \hat{\mathcal{L}}_\gamma}.
\end{equation}
Therefore, walker propagation reduces to applying exponentials of one-body operators and is relatively inexpensive compared to evaluating observables. Also, the decomposition of the bilinear coupling term results in separable and efficient propagation of electronic and photonic walker wavefunctions~\cite{Zhang:2025qmc}. The complexities of propagating electronic and photonic walker wavefunctions are $\mathcal{O}(N_\gamma NO)\propto \mathcal{O}(N^3)$ and $\mathcal{O}(P F^2)$ per walker, respectively, and $O$ is the number of occupied orbitals. Hence, the major computational bottleneck in AFQMC is the local energy estimation, as discussed later.

\subsection{Energy measurements and exchange energy in AFQMC}

The local energy (of walker $w$) is calculated as
\begin{align}\label{eq:etot}
    E^{w} = & \frac{\bra{\Psi_T}\hat{H}\ket{\psi_w}}{\bra{\Psi_T}\psi_w\rangle} = E_T + E_c + E_{e-ph} + E_{ph},
\end{align}
which is decomposed into kinetic, Coulomb, electron-photon interaction, and photonic energies:
\begin{align}
    E^w_T = & \sum_{pq}  h_{pq} \frac{\bra{\Psi_T} \hat{c}^\dagger_p \hat{c}_q  \ket{\psi_w}}{\bra{\Psi_T}\psi_w\rangle} = \sum_{pq} h_{pq} G^w_{pq},
    \\
    E^{w}_c = & \sum_{pqrs} V_{pqrs} \frac{\bra{\Psi_T}a^\dagger_p a^\dagger_q a_r a_s\ket{\psi_w}}{\bra{\Psi_T}\psi_w\rangle}
    \nonumber\\
    = & \sum_{pqrs} V_{pqrs} \left[G^w_{ps}G^{w}_{qr} - G^w_{pr}G^w_{qs} \right], \label{eq:coulomb_energy}
    \\
    E^w_{ph} =& \sum_{\alpha}\omega_\alpha  \frac{\bra{\Psi_T} \hat{a}^\dagger_\alpha  \hat{a}_\alpha  \ket{\psi_w}}{\bra{\Psi_T}\psi_w\rangle} = \sum_\alpha \omega_{\alpha} D^{w}_{\alpha},
    \\
   E^w_{e-ph} 
   = & \sum_{pq\alpha} \sqrt{\frac{\omega_\alpha}{2}} g^\alpha_{pq}  G^{w}_{pq} Q^\omega_\alpha.
\end{align}
Where  $\Psi_T$ and $\psi_w$ are the trial and walker wavefunctions, respectively. Generalized Wick's theorem is used to decompose the expectation value of a two-body operator into one-body terms. $G^w$ is the one-particle Green's function for the walker $w$ (or walker density matrix), which is given by
\[
   G^w_{pq} =  \frac{\bra{\Psi_T}a^\dag_p a_q\ket{\psi_w}}{\bra{\Psi_T}\psi_w\rangle} =
   \left[\psi_w (\Psi^\dagger_T \psi_w)^{-1}\Psi^\dagger_T\right]_{pq}.
\]
The complexity of computing kinetic, Coulomb, electron-photon interaction, and photonic energies is $\mathcal{O}(ON)$, $\mathcal{O}(O^2N^2)$, $\mathcal{O}(ONPF)$, and $\mathcal{O}(PF^2)$, respectively. Since the number of Fock states is small in general ($<10$), the number of modes needed in the calculation is also smaller than $ON$ in a molecular ensemble. The major bottleneck in the calculation is the Coulomb term that has scaling of  $\mathcal{O}(O^2N^2)$ or $\mathcal{O}(N^4)$ (since $O=\alpha N$). In particular, the exchange energy (second term in Eq.~\ref{eq:coulomb_energy})
\begin{equation}
    E^w_X =  \sum_{pqrs} V_{pqrs}  G^w_{pr}G^w_{qs}
    \label{eq:exchange}
\end{equation}
is the major bottleneck in energy measurement.
A brute-force computation scales as $\mathcal{O}(N^4)$. However, since $\Psi_T$ is fixed during the imaginary-time evolution (ITE) and has a size of $O\times N$, we can use $\Psi_T$ to transform the ERI into a new picture to reduce the memory cost. The resulting exchange energy becomes
\begin{equation}
    E_X = \sum_{pqij} \tilde{V}_{pqij} \Theta^w_{pi} \Theta^w_{qj}.
    \label{eq:ex_rotated}
\end{equation}
where
\[
  \Theta^w_{pi} = \left[\psi_w(\Psi^\dagger_T \psi_w)^{-1}\right]_{pi}
\]
is the half-rotated GF and $\tilde{V}_{pqij} = \sum_{rs} V_{pqrs}\Psi^\dagger_{T, ri}\Psi^\dagger_{T,sj}$ is the half-rotated ERI.
Consequently, the scaling of measuring exchange energy becomes $\mathcal{O}(O^2N^2)$ in both memory and wall time. Since the occupied orbitals are a small portion of the total AOs, particularly for a large basis set, the rotated picture leads to a significant reduction in the prefactor of the scaling. But the overall scaling is still quadratic. This quartic scaling makes AFQMC simulations prohibitive for large molecular systems and, especially, for ensembles of many molecules coupled to cavity photon modes. 

\subsection{Exchange Energy in Cholesky decomposition Formalism}
The CD method introduced before for rewriting the Hamiltonian into MC format can also be utilized to mitigate the $\mathcal{O}(N^4)$ scaling in storage. 
Substituting Eq.~\ref{eq:cd} into Eq.~\ref{eq:ex_rotated}, the exchange energy becomes
\begin{align}
    E_X &= \sum_{\gamma} \sum_{pqrs} L^\gamma_{pr} G_{ps} L^\gamma_{qr} G_{qs} \nonumber \\
    & = \sum_{\gamma, pqij}\tilde{L}^\gamma_{pi}\tilde{L}_{qj}\Theta_{pi}\Theta_{qj}
    = \sum_{\gamma, ij} f_{ij}^\gamma f_{ij}^\gamma,
    \label{eq:exchagne_cd}
\end{align}
where $f_{ij}^{\gamma} \equiv \sum_{r} \tilde{L}^{\gamma}_{ri}\, \Theta_{rj}$ and $\tilde{L}^{\gamma}_{pi} \equiv \sum_{q} L^{\gamma}_{pq}\, \Psi_{T,qi}$ are, respectively, the intermediate contraction tensors and the trial-rotated Cholesky tensor. From Eq.~\ref{eq:exchagne_cd}, it follows that the CD representation of the ERIs reduces the storage requirement to $\mathcal{O}(O N N_\gamma)$, compared with the explicit storage of the rotated four-index integrals $V_{pqij}$, which requires $\mathcal{O}(O^{2} N^{2})$. However, the computational scaling of the exchange energy in the CD form (Eq.~\ref{eq:exchagne_cd}) is $\mathcal{O}(O^{2} N N_\gamma)$, which is generally more expensive than Eq.~\ref{eq:ex_rotated} because in practice $N_\gamma > N$. Thus, the storage advantage of CD is accompanied by increased computational cost.

\subsection{Exchange Energy in tensor hypercontraction (THC) formalism}

To further reduce the memory cost and potentially reduce the computational scaling, THC was proposed for computing the exchange energy~\cite{Motta2019qmc}. THC can be regarded as a nested SVD and low rank approximation to the Cholesky tensor. That is, for a given $\gamma$, the $N \times N$ matrix $L^\gamma_{pq}$ can be approximated as~\cite{Motta2019qmc, Peng:2017uv}
\begin{equation}
    L^\gamma_{pq} \approx \sum_{\mu=1}^{N_\mu} X^\gamma_{p\mu} U^\gamma_{q\mu},
\end{equation}
where $N_\mu < N$ is the effective rank and is controlled by the threshold of the low rank approximation.
The THC representation compresses each Cholesky vector into two tall-and-skinny matrices $\{X^\gamma, U^\gamma\}$, reducing storage from $\mathcal{O}(N^2)$ to $\mathcal{O}(N N_\mu)$.

\begin{figure}[!htb]
    \centering
  \begin{tikzpicture}
    \node[inner sep=0] (imgA) {\includegraphics[width=.99\linewidth]{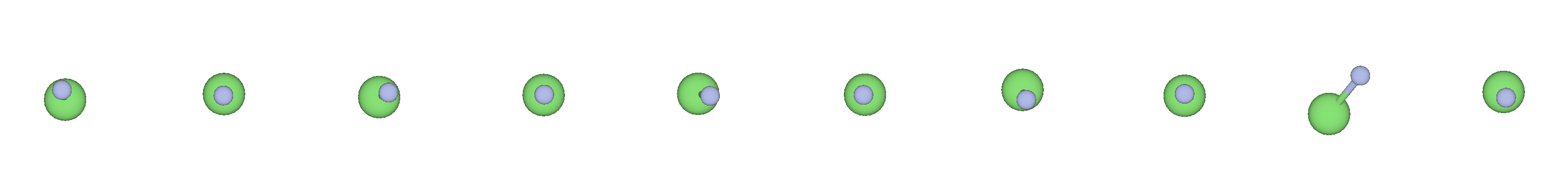}};
    \node[panel label, xshift=2pt, yshift=-2pt] at (imgA.north west) {a)};
    \draw[dashed, thick] ([yshift=5pt]imgA.south west) -- ([yshift=5pt]imgA.south east);
  \end{tikzpicture}
  \begin{tikzpicture}
    \node[inner sep=0] (imgA) {\includegraphics[width=.480\linewidth]{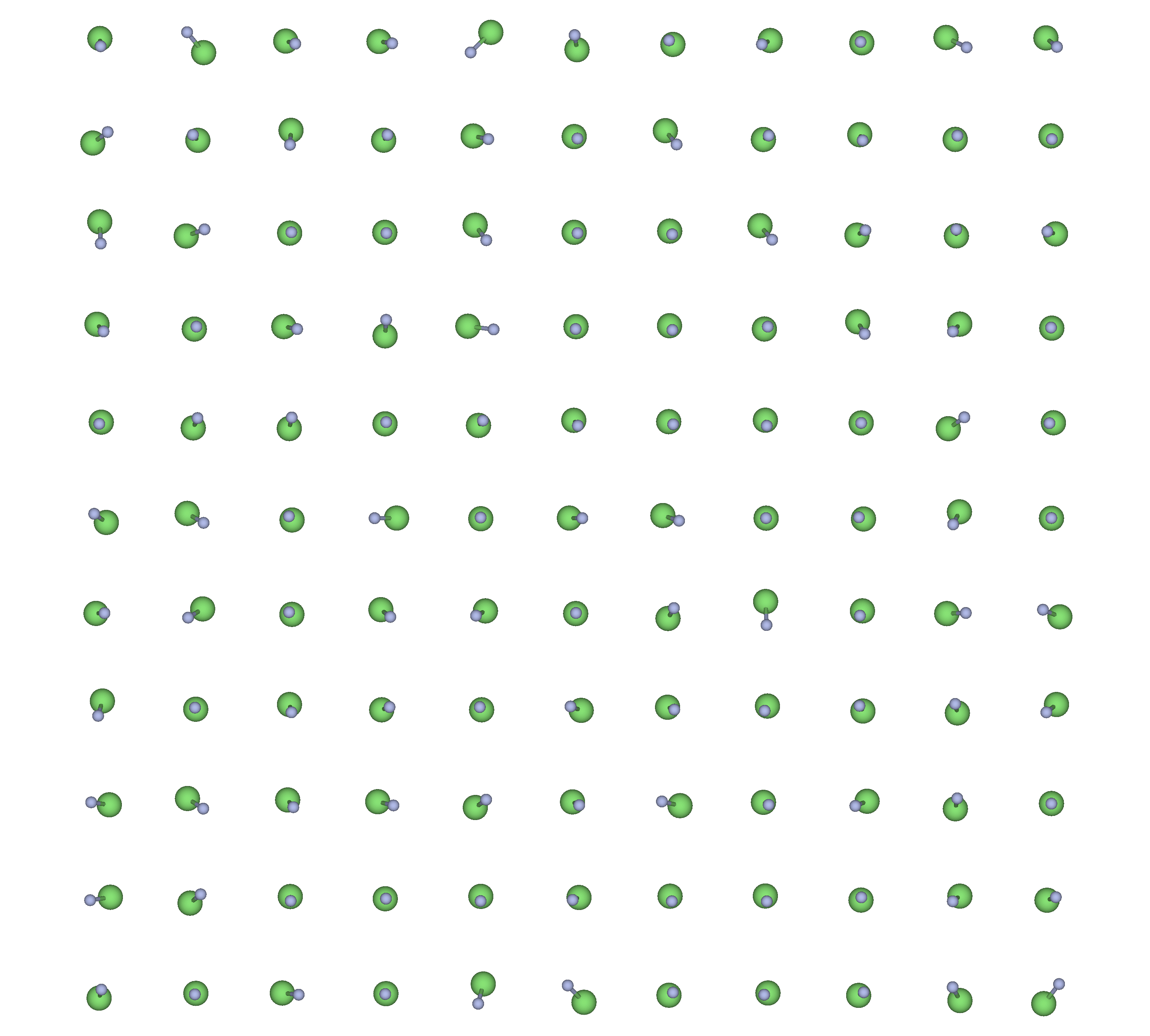}};
    \node[panel label, xshift=-1pt, yshift=0pt] at (imgA.north west) {b)};
    \draw[dashed, thick] ([yshift=0pt]imgA.north east) -- ([yshift=0pt]imgA.south east);
  \end{tikzpicture}
  \begin{tikzpicture}
    \node[inner sep=0] (imgA) {\includegraphics[width=.480\linewidth]{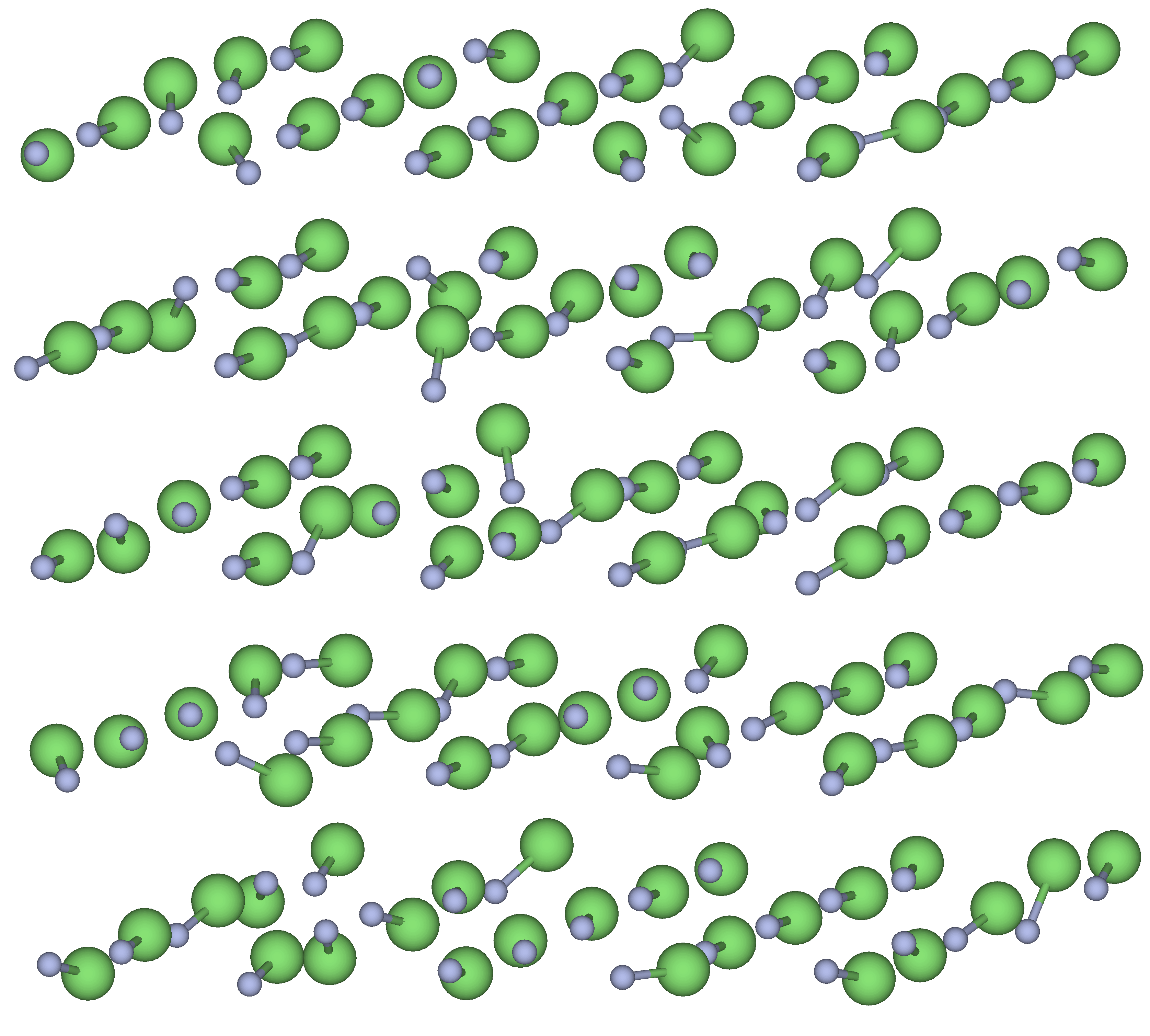}};
    \node[panel label, xshift=0pt, yshift=0pt] at (imgA.north west) {c)};
  \end{tikzpicture}
    \caption{Schematic 1D (a), 2D (b), and 3D (c) ensembles used in benchmarks. Random molecular orientations are used to avoid any symmetries.}
    \label{fig:molecules23d}
\end{figure}
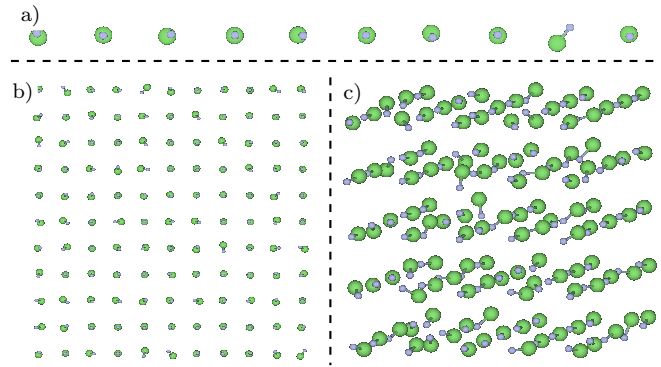

Inserting this factorization into the exchange energy yields
\begin{equation}\label{eq:fthc}
    f^\gamma_{ij} = \sum_{\mu=1}^{N_\mu} A^\gamma_{i\mu} B^\gamma_{j\mu},
\end{equation}
with $A^\gamma_{i\mu} = \sum_p T_{pi} X^\gamma_{p\mu}$ and $B^\gamma_{j\mu} = \sum_q U^\gamma_{q\mu} \Theta_{qj}$. 
Thus the cost of evaluating $f^\gamma_{ij}$ scales as $\mathcal{O}(O N N_\mu)$ per $\gamma$, giving a total cost of $\mathcal{O}(O^2 N_\gamma \langle N_\mu \rangle)$.
At fixed truncation accuracy (up to possible logarithmic factors), the effective rank $\langle R_\gamma \rangle$ was observed to saturate and become system-size independent only at very large sizes. Consequently, the scaling of using THC for exchange energy becomes effectively subquartic~\cite{Motta2019qmc}. However, as shown in early work, $\langle N_\mu \rangle$ only saturates in really large systems~\cite{}. For our molecular ensembles (shown in Figure~\ref{fig:molecules23d}), using a CD threshold of $10^{-4}$ for the ERIs, no saturation is observed with up to 1200 orbitals even in 1D (Figure~\ref{fig:averageranks}a). As a result, the method attains subquartic or cubic scaling only in the extreme asymptotic regime and carries a large prefactor at system sizes of practical interest.

\section{Efficient mixed block-sparsity and THC scheme for efficient calculation of exchange energy}
\label{sec:mixed}

\begin{figure}[!htb]
    \centering
  \begin{tikzpicture}
    \node[inner sep=0] (imgA) {\includegraphics[width=.485\linewidth]{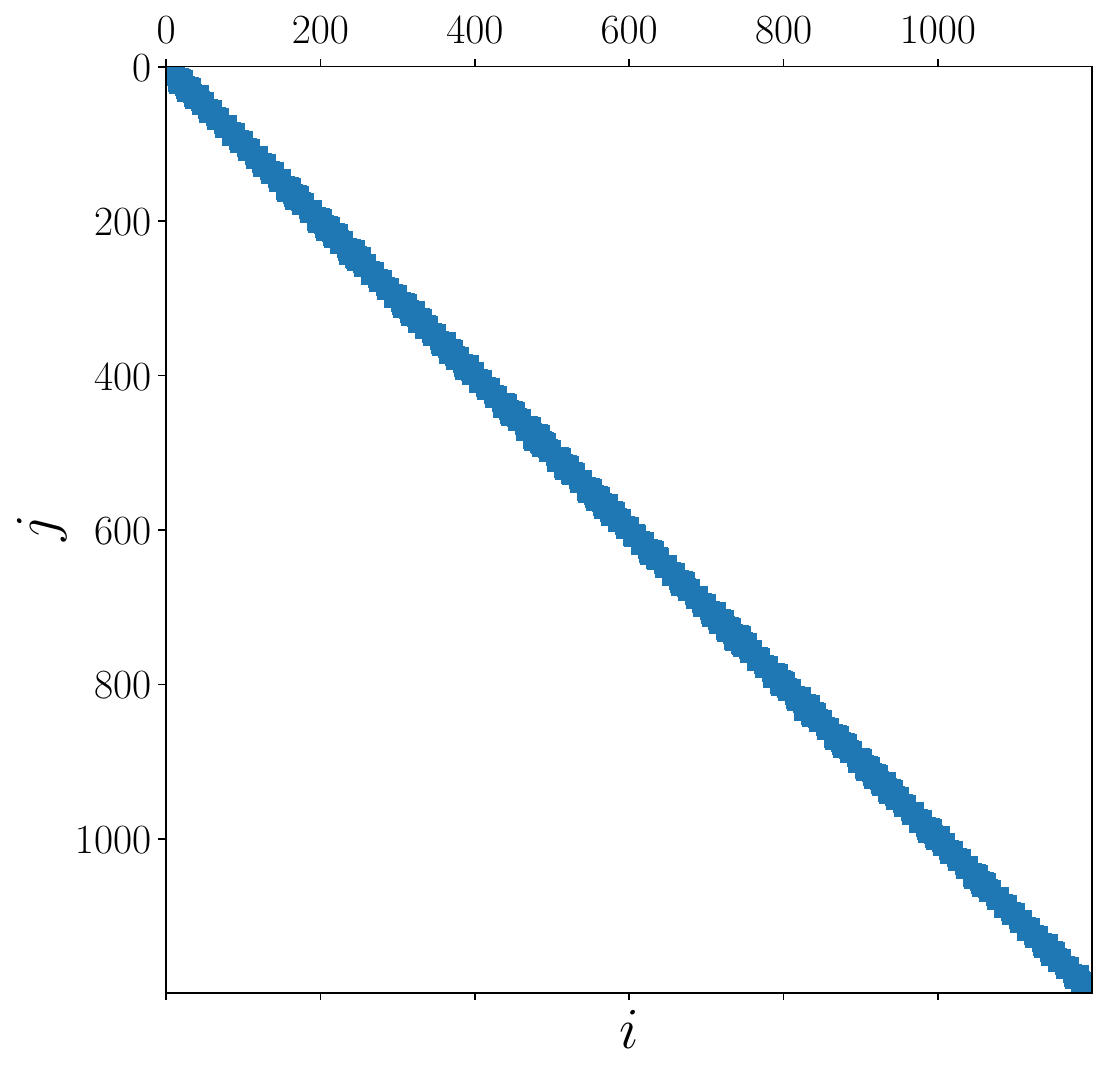}};
    \node[panel label, xshift=2pt, yshift=-2pt] at (imgA.north west) {a)};
  \end{tikzpicture}
    \begin{tikzpicture}
    \node[inner sep=0] (imgA) {\includegraphics[width=.495\linewidth]{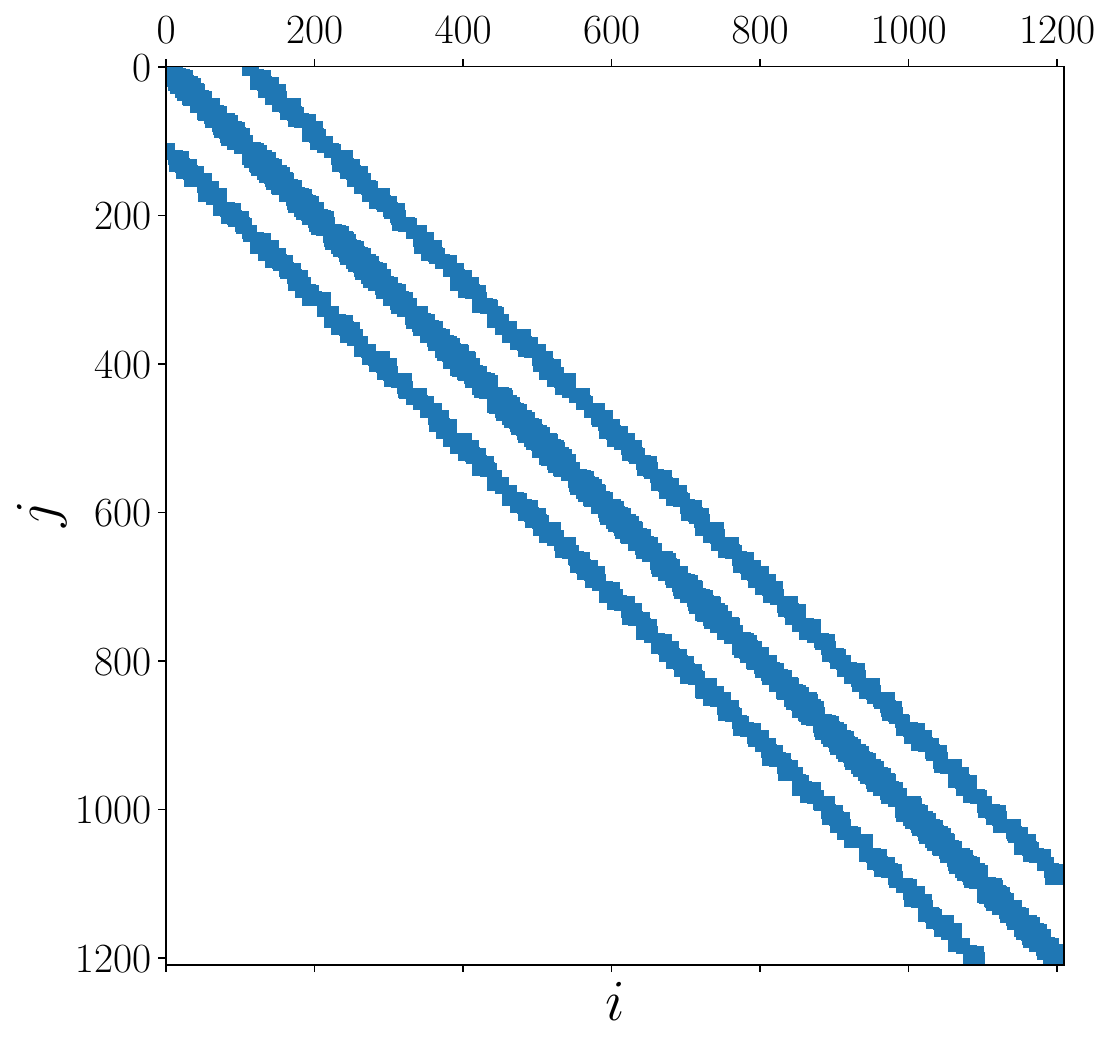}};
    \node[panel label, xshift=2pt, yshift=-2pt] at (imgA.north west) {b)};
  \end{tikzpicture}
   \begin{tikzpicture}
    \node[inner sep=0] (imgA) {\includegraphics[width=.5\linewidth]{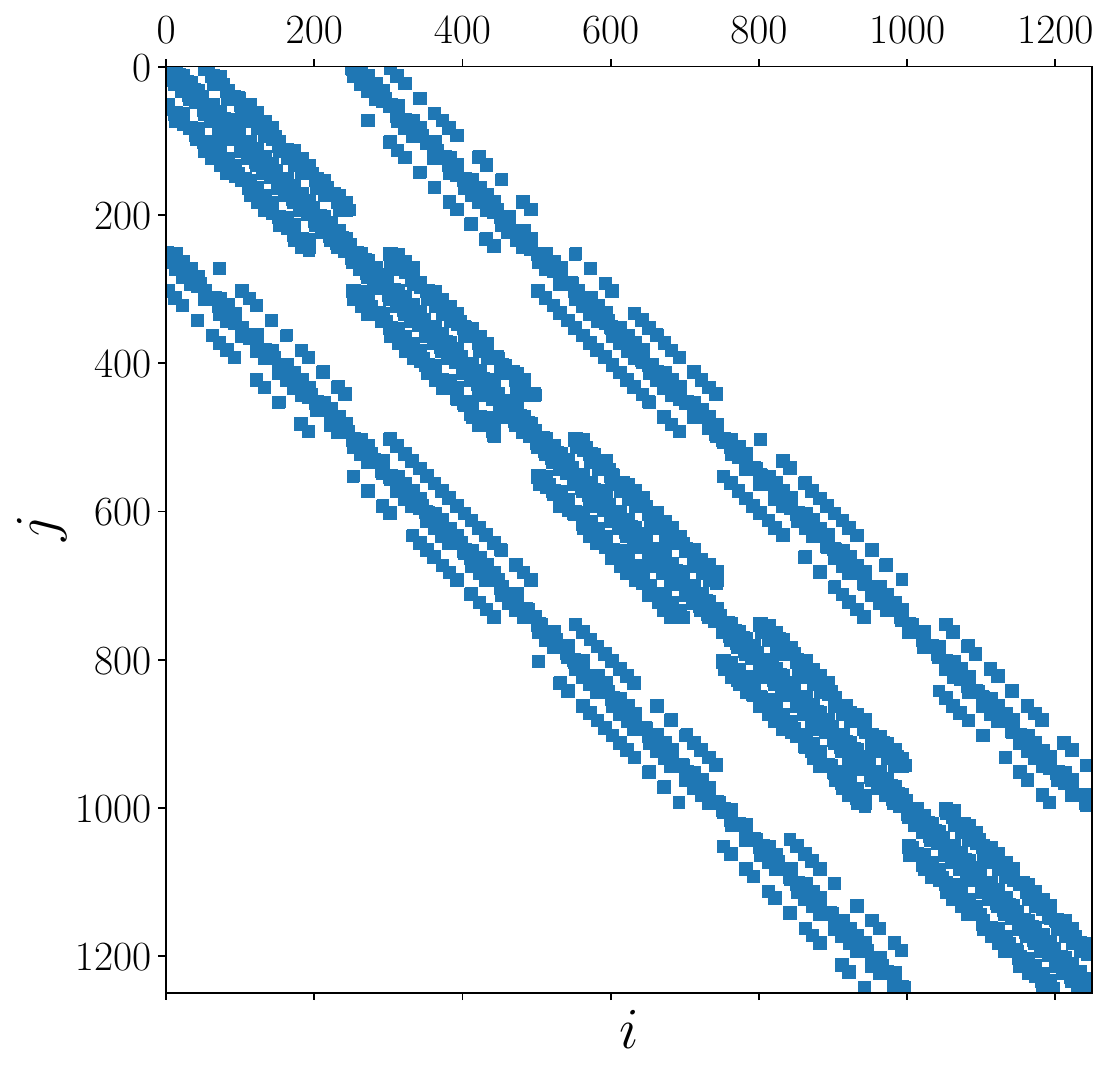}};
    \node[panel label, xshift=2pt, yshift=-2pt] at (imgA.north west) {c)};
  \end{tikzpicture}
    \begin{tikzpicture}
    \node[inner sep=0] (imgA) {\includegraphics[width=.48\linewidth]{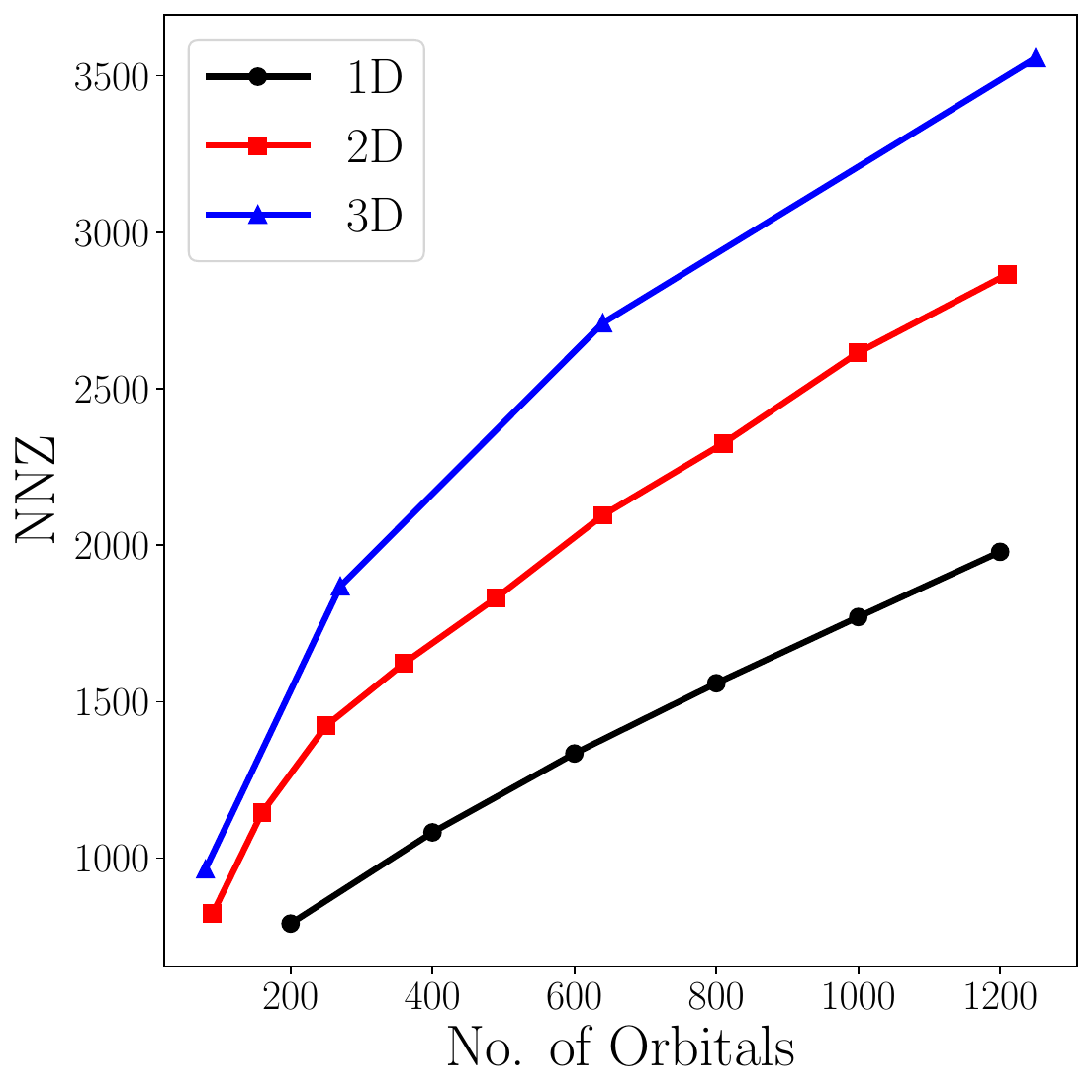}};
    \node[panel label, xshift=2pt, yshift=-2pt] at (imgA.north west) {d)};
  \end{tikzpicture}
    \caption{
   Sparsity of the averaged Cholesky tensor $\langle L\rangle$ for a) 1D (120 molecules), b) 2D ($11^2=121$ molecules), and c) 3D ($5^3=125$ molecules) ensembles: block diagonality with a narrow neighbor stencil is evident in all cases. d) Average NNZ per $L^\gamma$ versus $N$; the linear trend holds across dimensions with geometry-dependent slopes, supporting $\mathrm{NNZ}(L^\gamma)=\mathcal{O}(N)$.
    }
    \label{fig:sparsity}
\end{figure}
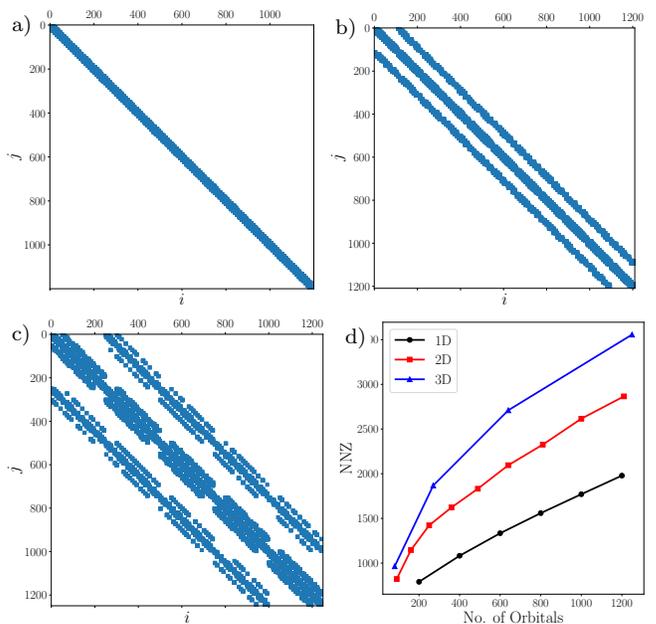

To address the bottleneck associated with THC, we develop a reduced-scaling strategy that exploits two key structural features of the Cholesky tensor in molecular ensembles: (i) the BS of the Cholesky tensor due to spatial locality and molecular separation, and (ii) the low rank nature of many Cholesky tensors, which can be efficiently factorized via THC.
As an example, we plotted the sparsity of the average of the Cholesky tensor $\langle L\rangle \equiv \frac{1}{N_\gamma}\sum_\gamma L_\gamma$ in Figure~\ref{fig:sparsity}a. The plot indicates that the Cholesky tensor shows block sparsity in either 1D, 2D, or 3D systems. Particularly, the Cholesky tensor is block tridiagonal in a 1D system. On the other hand, Cholesky also demonstrates inhomogeneous rank distribution as shown in Figure~\ref{fig:averageranks}b. Though there is a significant portion of the Cholesky tensor with low rank, the leading Cholesky (with larger norms) has a high rank (close to $N$), so that low rank approximation is not the optimal way to compress the data.

Based on the rank and sparsity analysis, the majority of Cholesky tensors are observed to have low ranks. For such blocks, THC provides a highly efficient compression. Only a minority of blocks—those with larger rank or less favorable conditioning—are more efficiently retained in their sparse CD form. 
We therefore adopt a \emph{mixed BS-THC} representation for the Cholesky tensors:
\begin{equation}
    \{ L^\gamma \} = \{ L^{\text{BS}} \} \cup \{ L^{\text{THC}} \},
\end{equation}
where $\{L^{\text{BS}}\}$ denotes the subset of Cholesky tensors stored in block-sparse format, and $\{L^{\text{THC}}\}$ denotes the low rank subset represented by THC. This mixed BS-THC decomposition balances the strengths of both schemes:
Block sparsity minimizes operations on large-rank vectors, while THC efficiently compresses low rank ones. As shown in the following, the overall computational scaling for exchange energy evaluation in the mixed BS-THC scheme becomes $\sim \mathcal{O}(N^3)$ with a reduced memory footprint compared to pure CD or THC approaches.

Let $N_b$ denote the number of nonzero diagonal blocks and $s\equiv N/N_b$ the block size.
Let $d$ be the average block degree (number of neighbor blocks; which is $\mathcal{O}(1)$ in the 1D/2D/3D setups considered). For instance, $d=2$ for the 1D system, i.e., $L_\gamma$ is block tridiagonal for the 1D system as shown in Figure~\ref{fig:sparsity}. 
Hence, within the BS format, each $L^\gamma$ contains on the order of $(d+1)N_b$ dense $s\times s$ blocks and therefore the total number of nonzeros (NNZ) in each $L^\gamma$ is
\begin{equation}
  \mathrm{NNZ}(L^\gamma) \approx (d+1) N_b s^2  = (d+1) N s  = \mathcal{O}(N),
  \label{eq:linearnnz}
\end{equation}
with a slightly larger block degree ($d$) in higher dimensions due to increased adjacency.
With localized $\Psi_T$ (we use molecule-localized occupied orbitals, so the half-rotated $\tilde{L}^\gamma=L^\gamma \Psi_T$ inherits the block pattern up to a constant-width stencil), the per-$\gamma$ cost to compute $f_{ij}^\gamma$ in the BS format is
\begin{equation}
    \mathcal{C}_{\mathrm{BS}}(\gamma) \sim c_{\mathrm{BS}} (d+1)s ON \propto N^2,
    \label{eq:cost_bs}
\end{equation}
since $O\propto N$ at fixed filling and $(d{+}1)s$ is independent of $N$.
In contrast, the per-$\gamma$ cost in the THC formalism (Eq.~\ref{eq:fthc}) with rank $R_\gamma$ is,
\begin{equation}
    \mathcal{C}_{\mathrm{THC}}(\gamma) \sim c_{\mathrm{THC}} N O R_\gamma 
    \propto N^2 R_\gamma.
    \label{eq:cost_thc}
\end{equation}
Equating Eqs.~\eqref{eq:cost_bs} and \eqref{eq:cost_thc} yields an \emph{optimal, size-independent} rank threshold that determines which scheme is more efficient for computing the exchange energy,
\begin{equation}
    R_\gamma^\star \approx \kappa (d{+}1)s, 
    \label{eq:rank_star}
\end{equation}
with $\kappa \equiv c_{\mathrm{BS}}/c_{\mathrm{THC}}$ absorbing constants. Thus, the decision rule for determining the optimal cut between the BS and THC is
\begin{align}
    L^\gamma \in 
    \begin{cases}
      \text{THC} & \text{if } R_\gamma \le R_\gamma^\star,\\[2pt]
      \text{BS}  & \text{if } R_\gamma > R_\gamma^\star.
    \end{cases}
    \label{eq:decision_rule}
\end{align}
Because $R_\gamma^\star$ is $\mathcal{O}(1)$ with respect to $N$, the THC set contains only genuinely low rank vectors, preventing super-cubic growth.  
The threshold $R_\gamma^\star$ in Eq.~\eqref{eq:rank_star} is obtained by balancing the leading per-$\gamma$ costs of the BS and THC evaluation and should be viewed as a crossover estimate rather than a sensitive tuning parameter. Moderate changes in $R_\gamma^\star$ primarily reassign a relatively small set of intermediate-rank Cholesky tensors between the two subsets; genuinely low-rank tensors remain efficiently handled by THC while high-rank tensors remain in the BS set. As a result, the overall $\mathcal{O}(N^3)$ scaling is unchanged, and for fixed THC truncation accuracy the total energy is insensitive to the precise value of $R_\gamma^\star$; varying the threshold mainly affects computational prefactors (wall time and memory) rather than accuracy.

A BS format of $L^\gamma$ needs $\mathcal{O}(\mathrm{NNZ}(L^\gamma))=\mathcal{O}(N)$ storage. A THC set of $L^\gamma$ needs $\mathcal{O}(N R^*_\gamma)\propto\mathcal{O}(N)$. Summed over $N_\gamma=\mathcal{O}(N)$ Cholesky tensors, the memory of the mixed BS-THC scheme scales as $\mathcal{O}(N^2)$, whereas pure CD stores $\mathcal{O}(O N N_\gamma)=\mathcal{O}(N^3)$ floating-point numbers.
With $N_\gamma\propto N$ and $R^*_\gamma$ in THC capped by Eq.~\eqref{eq:rank_star}, the total work over all $\gamma$ is
\[
  \sum_{\gamma\in\mathrm{BS}} \mathcal{C}_{\mathrm{BS}}(\gamma) + 
  \sum_{\gamma\in\mathrm{THC}} \mathcal{C}_{\mathrm{THC}}(\gamma)
  \ =\ \mathcal{O}(N^3).
\]
This cubic scaling is robust across 1D/2D/3D geometries since $d$ and $s$ are $N$-independent.

Finally, it should be noted that the mixed BS--THC representation is constructed once as a preprocessing step after the (modified) Cholesky decomposition of the ERI and before AFQMC propagation. In practice, building the BS/THC factors introduces only a small additional overhead relative to the CD step. The resulting factors are then reused at every propagation step, so the one-time setup cost is negligible compared to the total AFQMC runtime. For clarity and to facilitate independent implementation, Algorithm~\ref{alg:mixed_bsthc} summarizes the complete workflow: the one-time construction of the mixed BS--THC representation from Cholesky tensors, followed by the per-walker exchange-energy evaluation used in AFQMC measurements.

\begin{algorithm}[H]
\caption{Mixed BS--THC workflow for exchange-energy evaluation in AFQMC}
\label{alg:mixed_bsthc}
\begin{algorithmic}[1]
\Require DSE-modified integral constructor $(\tilde{h},\tilde{V})$, trial determinant $\Psi_T$, low-rank threshold $\tau_{\mathrm{CD}}$, THC tolerance $\tau_{\mathrm{THC}}$, element truncation $\eta$, 
walkers $\{\psi_w\}$, rank cutoff $R^\star$.

\vspace{2pt}
\Statex \textbf{(A) One-shot preprocessing: build mixed BS--THC factors}
\State Compute Cholesky tensors $\{L^\gamma\}_{\gamma=1}^{N_\gamma}$ from sliced $\tilde{V}$ constructor (instead of explicitly computing entire $\tilde{V}$ to save memory) using CD with threshold $\tau_{\mathrm{CD}}$.
\State Initialize $\{L^{\mathrm{BS}}\}\gets \emptyset$, $\{L^{\mathrm{THC}}\}\gets \emptyset$.
\For{$\gamma=1,\dots,N_\gamma$}
    \State Truncate small entries: $L^\gamma_{pq}\leftarrow 0$ if $|L^\gamma_{pq}|<\eta$.
    \State Estimate numerical rank $R_\gamma$ of $L^\gamma$ at tolerance $\tau_{\mathrm{THC}}$ via $L^\gamma_{pq} \approx \sum_{\mu=1}^{R_\gamma} X^\gamma_{p\mu}U^\gamma_{q\mu}$.
    \If{$R_\gamma \le R^\star$}
        \State Compute trial-rotated $X^\gamma$: $A^\gamma \gets \Psi_T^\dagger X^\gamma$ and Store $(A^\gamma,U^\gamma)$ in $\{L^{\mathrm{THC}}\}$.
    \Else
        \State Store $L^\gamma$ in block-sparse (BSR-like) format 
        \State Form trial-rotated tensor $\tilde{L}^\gamma \gets L^\gamma \Psi_T$ 
         and append to $\{L^{\mathrm{BS}}\}$.
    \EndIf
\EndFor

\vspace{2pt}
\Statex \textbf{(B) Exchange energy measurement during the AFQMC propagation}
\For{each walker $\psi_w$}
    \State Compute overlap  and $\Theta \gets \psi_w (\Psi_T^\dagger \psi_w)^{-1}$.
    \State Initialize $E_X^w \gets 0$.
    \For{each $\tilde{L}^\gamma \in \{L^{\mathrm{BS}}\}$}
        \State Compute $F^\gamma \gets (\tilde{L}^\gamma)^\dagger \Theta$.
        \State Accumulate $E_X^w \gets E_X^w + \|F^\gamma\|_F^2$.
    \EndFor
    \For{each $(A^\gamma,U^\gamma) \in \{L^{\mathrm{THC}}\}$}
        \State Compute $B^\gamma \gets (U^\gamma)^{\mathsf T}\Theta$.
        \State Form $F^\gamma \gets A^\gamma (B^\gamma)^{\mathsf T}$.
        \State Accumulate $E_X^w \gets E_X^w + \|F^\gamma\|_F^2$.
    \EndFor
\EndFor
\State \Return $\{E_X^w\}$.
\end{algorithmic}
\end{algorithm}

\section{Results and Discussion}
\label{sec:results}

We use the diatomic LiF molecule as a representative monomer to demonstrate the proposed methodology. Molecular ensembles are constructed by replicating the monomer along one, two, and three spatial directions, with random molecular orientations, to form 1D, 2D, and 3D arrays (Fig.~\ref{fig:molecules23d}). Representative sizes include 120 molecules in 1D chains, $11\times 11=121$ molecules in a 2D lattice, and $5^3=125$ molecules in 3D lattices, yielding $\mathcal{O}(10^3)$ orbitals in a minimal STO\mbox{-}3G basis.
The precise orbital counts depend on the basis and frozen-core choices, but the qualitative trends reported here are insensitive to these details. 
The CD tensors are generated using the modified Cholesky decomposition without explicitly forming the full four-index electron-repulsion integral (ERI) tensor. Unless otherwise stated, identical thresholds in CD and THC (i.e., $\tau_{\mathrm{CD}} = \tau_{\mathrm{THC}}$) are used across all geometries to enable fair comparisons. 

Figure~\ref{fig:molecules23d}a)-c) illustrate the molecular ensemble in 1D, 2D, and 3D simple-cubic ensemble layouts used throughout. Random molecular orientations remove artificial symmetry in the ERIs while preserving the short-range, block-local nature of the intra-molecular Coulomb term that underlies the sparsity we exploit. Throughout the manuscript, a block size of 20 is used. 

Figures~\ref{fig:sparsity}a)-c) plot the sparsity of the \emph{averaged} Cholesky tensor, $\langle L\rangle \equiv \frac{1}{N_\gamma}\sum^{N_\gamma}_\gamma L^\gamma$, reshaped as a 2D matrix for visualization.  A threshold of $\eta=10^{-6}$ is employed to truncate a matrix element as zero if its absolute value is smaller than $\eta$.
For all geometries in different dimensions, we observe pronounced block diagonality with a narrow band of neighboring blocks.
With the atomic orbitals ordered by molecule, we set the block size $s$ to twice the number of orbitals on a single molecule.
Figures~\ref{fig:sparsity}a)-c) demonstrate that the 1D system approaches a block tridiagonal pattern; 2D/3D introduce a few additional neighbor shells but maintain a constant block degree $d$ (diagonal $+$ a fixed number of neighbors) that is \emph{independent of system size}.
Figure~\ref{fig:sparsity}d quantifies Eq.~\ref{eq:linearnnz}, showing that the total NNZ per $L^\gamma$ grows asymptotically \emph{linearly} with the number of orbitals $N$ in 1D/2D/3D. The slopes differ modestly, reflecting geometry-dependent neighbor counts, but the \emph{topology} in BS is the same. This directly supports the $\mathcal{O}(N)$ memory complexity for each $L^\gamma$. Hence, the overall complexity of storing all the $\{L^\gamma\}$ tensors is reduced to $\mathcal{O}(N_\gamma N)\propto \mathcal{O}(N^2)$.

Using small blocks with fixed block size is significantly more memory efficient than enforcing a global \emph{block-tridiagonal} layout based on geometric layers. For example, in a $L\times L$ 2D array, a layer has a block size $s_\mathrm{layer}\sim L\,s$ and $N_b\sim L$, giving
$\mathrm{NNZ}(L^\gamma)\sim 3\,L\,(L s)^2=\mathcal{O}(N^{3/2})$; in 3D, with $L\times L\times L$ molecules, $s_\mathrm{layer}\sim L^2 s$ and $N_b\sim L$, so
$\mathrm{NNZ}(L^\gamma)\sim 3\,L\,(L^2 s)^2=\mathcal{O}(N^{5/3})$. In contrast,  using the BS format with a fixed block size $s$ preserves the desirable linear scaling $\mathcal{O}(N)$ per $L^\gamma$  in all dimensions. In addition, from an implementation standpoint, we store the BS format of $L^\gamma$ tensors in a block-based layout (e.g., Block Compressed Row (BSR)-like) rather than fine-grained Compressed Sparse Row (CSR). While this block format can use more memory than CSR (since each nonzero block is stored as a dense $s\times s$ matrix), it dramatically improves arithmetic intensity and enables the dominant contractions, such as $(\tilde{L}^\gamma)^{\dagger}\Theta$, to be executed as \emph{batched} matrix-matrix multiplications. 
This avoids pointer chasing and irregular gather/scatter, maps cleanly onto CPU cache hierarchies, and leverages high-throughput GPU libraries (e.g., cuBLAS/rocBLAS) on heterogeneous nodes. 

\begin{figure}[!htb]
  \begin{tikzpicture}
    \node[inner sep=0] (imgA) {\includegraphics[width=.48\linewidth]{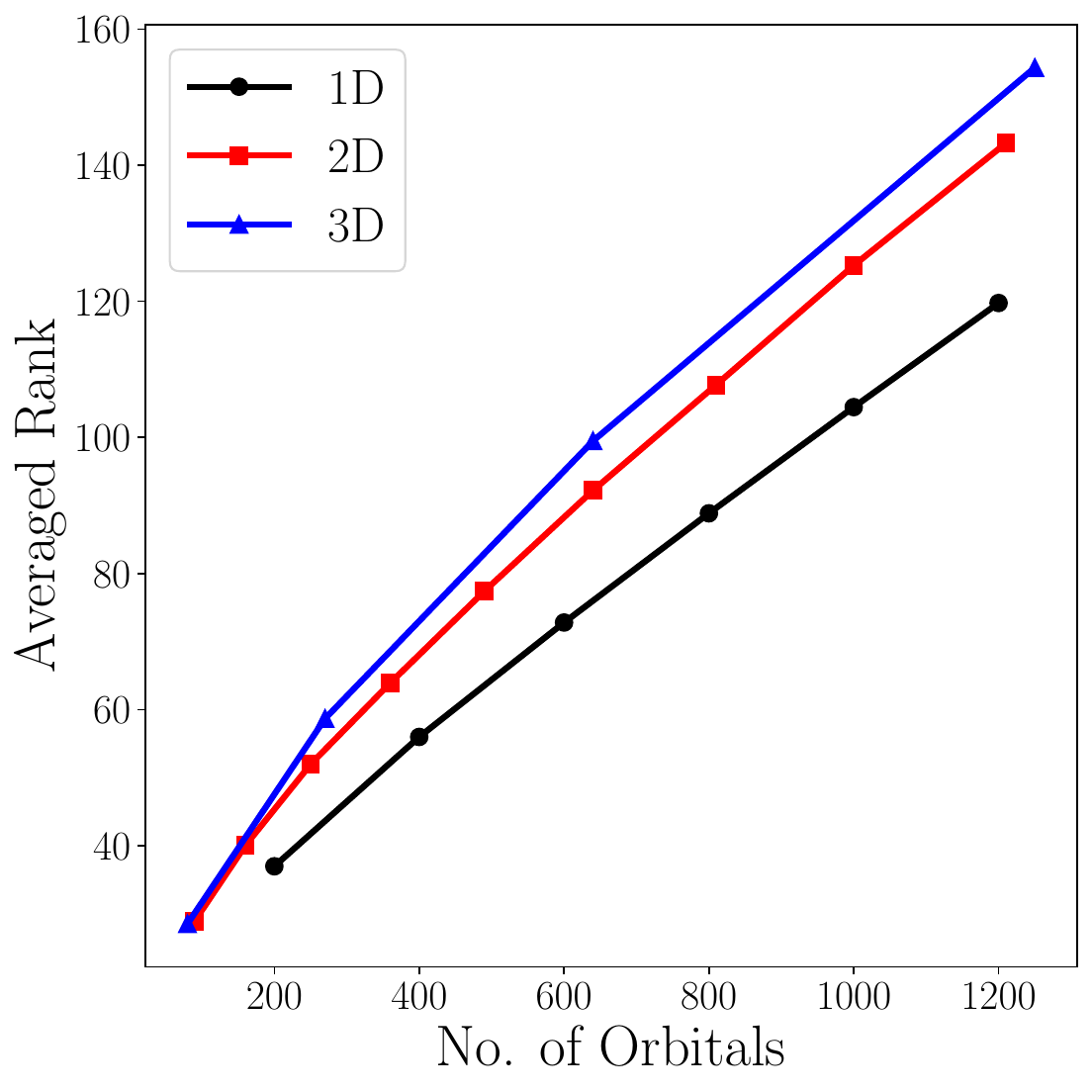}};
    \node[panel label, xshift=-2pt, yshift=-2pt] at (imgA.north west) {a)};
  \end{tikzpicture}
  \begin{tikzpicture}
    \node[inner sep=0] (imgA) {\includegraphics[width=.48\linewidth]{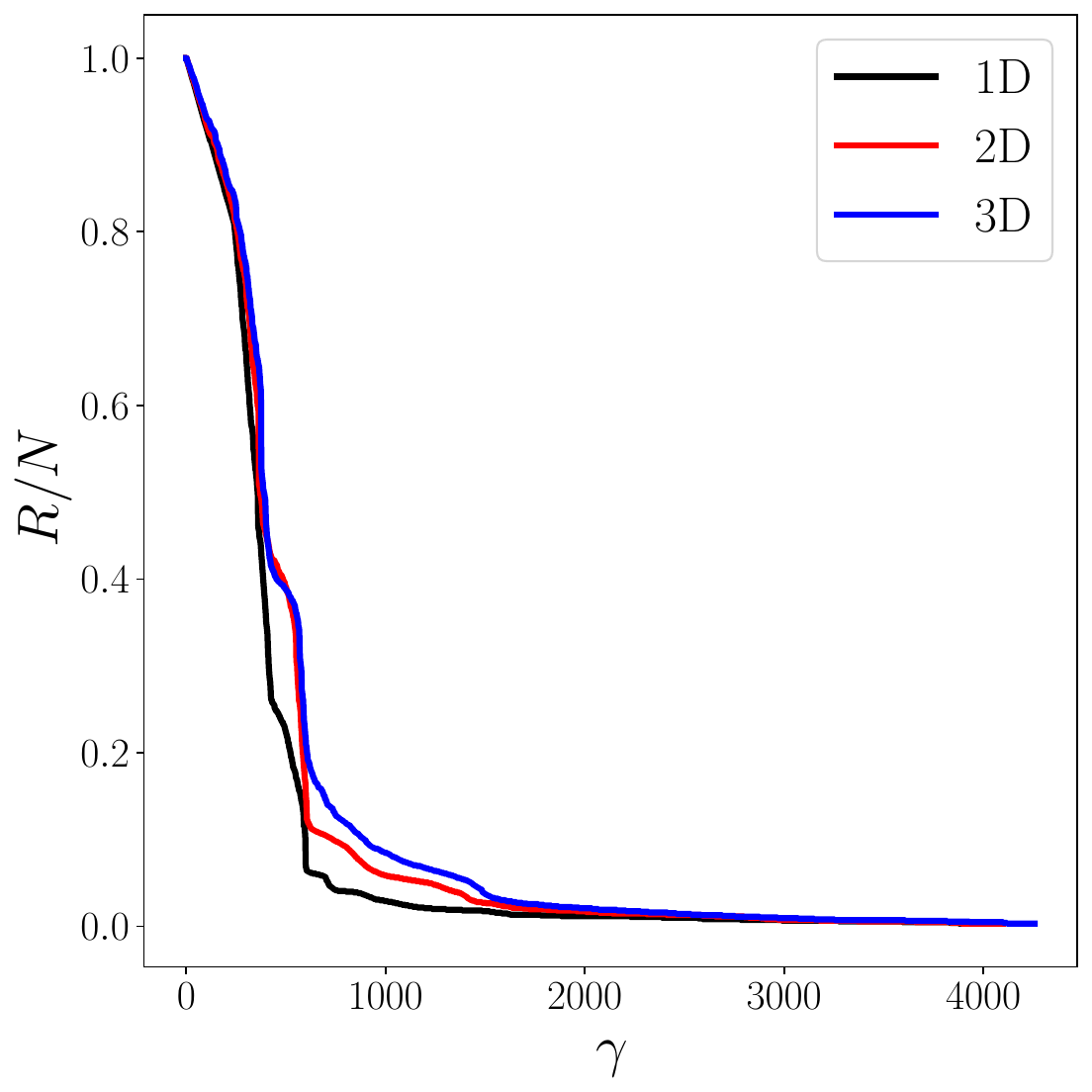}};
    \node[panel label, xshift=-2pt, yshift=-2pt] at (imgA.north west) {b)};
  \end{tikzpicture}
    \caption{
    (a) Average numerical rank $\bar{R}$ of $L^\gamma$ versus $N$ for 1D/2D/3D ensembles at fixed tolerance: $\bar{R}$ grows sublinearly and does not saturate up to $\sim$1200 orbitals, implying super-cubic/sub-quartic scaling for \emph{pure} THC. (b) Rank $R_\gamma$ versus Cholesky index: many genuinely low rank vectors coexist with a subset that is near full rank, motivating the split of Cholesky tensors into BS and THC sets in Eq.~\eqref{eq:decision_rule}. $\tau_{\mathrm{CD}} = \tau_{\mathrm{THC}}=10^{-4}$ is used.
    }
    \label{fig:averageranks}
\end{figure}

Figure~\ref{fig:averageranks}a) reports the average numerical rank $\bar{R}\equiv \frac{1}{N_\gamma} \sum R_\gamma$ obtained from a truncated SVD at fixed tolerance $\tau_{\mathrm{THC}}$, defined by the Frobenius residual
$\|L^\gamma - L^\gamma_{(R)}\|_F / \|L^\gamma\|_F \le \tau_{\mathrm{THC}}$.
Across 1D/2D/3D geometries, $\bar{R}$ grows \emph{sublinearly} with $N$ and does not saturate up to $\sim\!10^3$ orbitals. The trend is well described by a power law $\bar{R}\sim N^{\alpha}$ with $0<\alpha<1$ (dimension-dependent prefactors reflect different neighbor stencils). Consequently, a \emph{pure} THC strategy is super-cubic: the per-$\gamma$ cost scales as $\mathcal{O}(N^2 R_\gamma)$, and summing over $N_\gamma\propto N$ gives a total cost $\mathcal{O}\!\left(N^3 \bar{R}\right)=\mathcal{O}\!\left(N^{3+\alpha}\right)$ at fixed $\tau_{\mathrm{THC}}$. The same rank growth also inflates memory in pure THC to $\mathcal{O}\!\left(N^2\bar{R}\right)$, in contrast to the $\mathcal{O}(N^2)$ memory footprint achieved by our mixed BS-THC scheme.

\begin{figure}[!htb]
    \centering
    \includegraphics[width=0.96\linewidth]{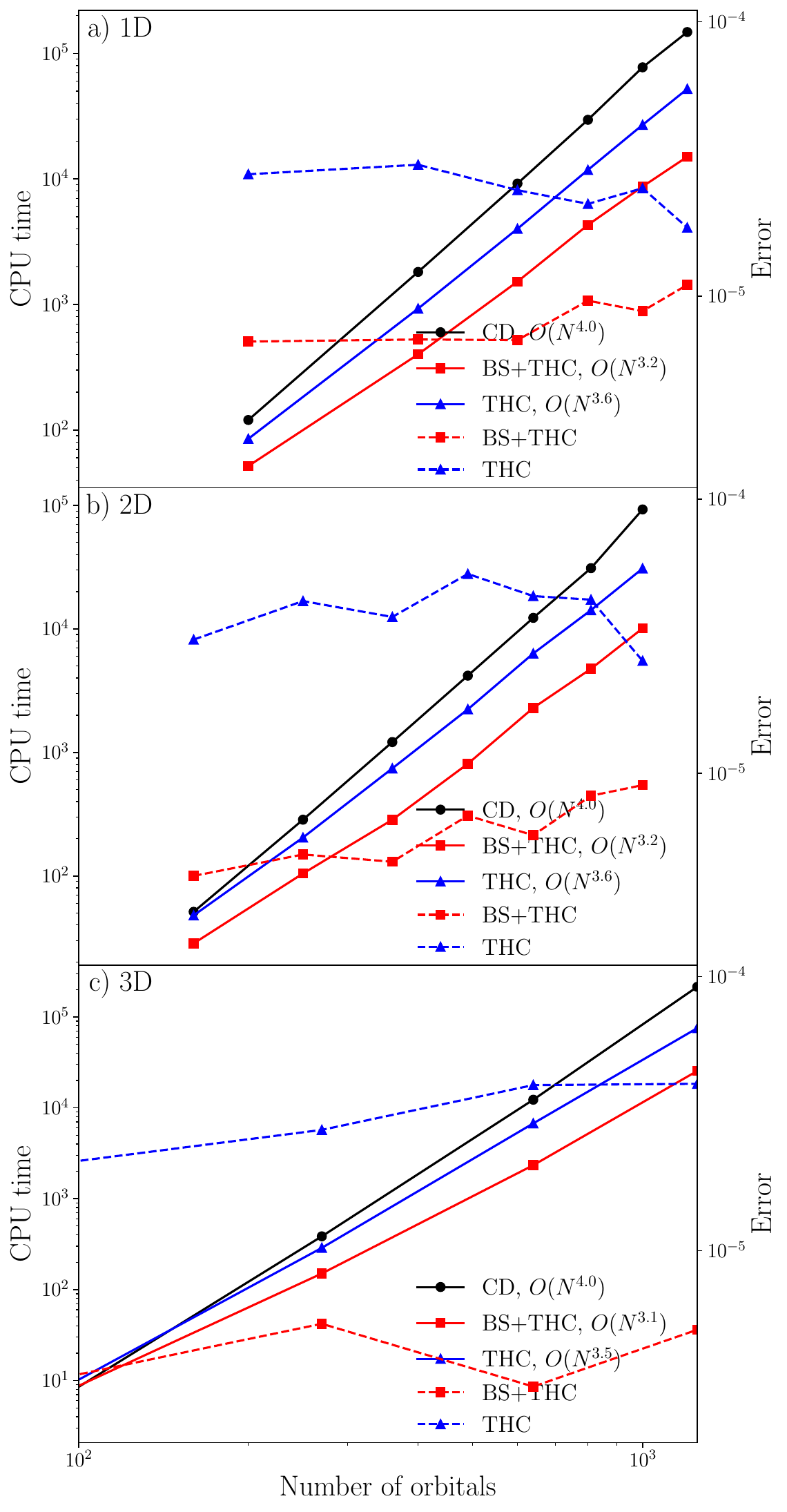}
    \caption{
    CPU-time scaling and accuracy of exchange-energy evaluation versus number of orbitals for (i) pure CD (quartic scaling; black), (ii) pure THC (super-cubic to sub-quartic due to rank growth; blue), and (iii) the proposed mixed BS–THC method (robust cubic scaling; red), shown for (a) 1D, (b) 2D, and (c) 3D molecular ensembles. The fitted scaling is included in the legend. Overall, the mixed scheme consistently achieves $\sim\mathcal{O}(N^3)$ scaling while preserving accuracy.
    }
    \label{fig:walltime}
\end{figure}

Figure~\ref{fig:averageranks}b) plots the individual ranks $R_\gamma$ versus the Cholesky index $\gamma$ and reveals pronounced \emph{heterogeneity}: many $L^\gamma$ are genuinely low rank, while a non-negligible subset is near full rank. Empirically, high-rank $L^\gamma$ correlate with larger $\|L^\gamma\|_F$ and capture short-range, intra-molecular Coulomb structure; the low rank tail primarily reflects smoother, longer-range contributions. This heterogeneity is precisely what the mixed BS-THC scheme leverages: we route the high-rank subset through the block-sparse (BS) path and compress only the genuinely low rank subset with THC. The decision rule in Eq.~\eqref{eq:decision_rule} uses a \emph{size-independent} threshold $R_\gamma^\star$ [Eq.~\ref{eq:rank_star}], derived by equating per-$\gamma$ costs. Keeping $R_\gamma \le R_\gamma^\star$ for the THC subset prevents $\bar{R}$ from feeding into the asymptotic exponent, so the overall scaling remains robustly cubic. In practice, the fraction of vectors routed to THC grows slowly with $N$ but remains bounded by the fixed threshold, while the BS path continues to benefit from $\mathrm{NNZ}(L^\gamma)=\mathcal{O}(N)$ due to the constant block degree.

Figure~\ref{fig:walltime} reports the measured CPU times and accuracy for exchange energy evaluation versus $N$ (in different dimensions) for three strategies: (i) pure CD (asymptotically $\mathcal{O}(N^{4})$), (ii) pure THC (super-cubic/sub-quartic due to the growth of the effective rank $\bar{R}$ at fixed tolerance), and (iii) the proposed mixed BS-THC scheme (cubic). 
Here, we report the exchange-only error, $\Delta E_x \equiv |\frac{E_x^{\mathrm{approx}} - E_x^{\mathrm{CD}}}{E_x^{\mathrm{CD}}}|$, where $E_x^{\mathrm{CD}}$ uses the CD Hamiltonian (same CD threshold) and $E_x^{\mathrm{approx}}$ additionally applies THC or mixed BS--THC to $\{L^\gamma\}$. 
Timings correspond to the exchange kernel only (setup costs for CD/THC factorizations are amortized) and were obtained with identical CD and THC thresholds across geometries to ensure a fair comparison. Power-law fits on a log-log scale yield slopes consistent with the expected asymptotics: the mixed BS-THC scheme exhibits $\beta_{\mathrm{hyb}}\approx 3$ within the fit uncertainty across 1D/2D/3D, while THC shows $\beta_{\mathrm{THC}}=3+\alpha$ with $0<\alpha<1$ reflecting $\bar{R}\sim N^{\alpha}$ (cf.\ Fig.~\ref{fig:averageranks}a), and CD approaches $\beta_{\mathrm{CD}}\approx 4$. Since a threshold of $10^{-6}$ is used to truncate the elements in the $L_\gamma$ tensor, the accuracy of the mixed BS-THC method surpasses that of the standard THC scheme. The efficiency of the mixed BS-THC scheme is expected to further improve with a larger threshold $\eta$. Consequently, Figure~\ref{fig:walltime} demonstrates that the mixed BS-THC scheme significantly improves computational efficiency without sacrificing accuracy, exhibiting consistent performance gains across 1D, 2D, and 3D systems.

Memory usage mirrors the timing trends. Pure CD requires storing $O\times N\times N_\gamma=\mathcal{O}(N^{3})$ rotated intermediates and frequently becomes memory bound at larger $N$. Pure THC reduces storage per $L^\gamma$ to $\mathcal{O}(NR_\gamma)$ but inherits the same $\bar{R}(N)$ growth, yielding $\mathcal{O}(N^{2}\bar{R})$ overall. The mixed BS-THC scheme keeps high-rank vectors in block-sparse blocks with $\mathrm{NNZ}(L^\gamma)=\mathcal{O}(N)$ and compresses only the low rank tail, resulting in an overall $\mathcal{O}(N^{2})$ footprint.
We note that the observed prefactor differences among 1D/2D/3D in Fig.~\ref{fig:walltime} track the average block degree $d$ (i.e., the number of neighbor blocks). The slopes remain geometry invariant because $d$ does not grow with $N$, but 2D/3D exhibit slightly larger constants than 1D. The mixed BS-THC scheme's combination of (i) constant-stencil block sparsity and (ii) bounded-rank THC therefore delivers robust $\mathcal{O}(N^{3})$ scaling with favorable prefactors and preserves accuracy at fixed thresholds, outperforming pure CD and pure THC well before the largest system sizes considered here.

While we used weakly interacting LiF molecular ensembles in a minimal basis primarily as a clean and scalable benchmark, the mixed BS–THC framework is not restricted to this setting and is expected to apply to more strongly interacting systems and larger basis sets, provided a reasonably localized representation can be constructed. To illustrate robustness beyond the LiF case (and minimal basis set), we additionally consider a C$_2$N$_2$H$_6$ molecular ensemble with the cc-pVDZ basis set at shorter intermolecular separation (which increases intermolecular coupling and delocalization). Despite the stronger interactions and the use of a larger basis, the Cholesky tensors still exhibit a block-sparse structure and pronounced rank heterogeneity (Figure~S1 in the Supporting Information), albeit with a larger effective block size and correspondingly larger prefactors. Applying the same mixed BS–THC partitioning and cost model, we still observe asymptotic cubic scaling for the exchange-energy evaluation.

In this work, we do not apply an explicit orbital-localization procedure (e.g., Boys or Pipek--Mezey~\cite{Foster:1960aa, Hoyvik:2012aa}) to enforce locality. Instead, the block sparsity observed in Fig.~\ref{fig:sparsity} primarily reflects the physical structure of the systems studied: weakly interacting molecular ensembles with appreciable spatial separation. For more strongly interacting complexes or extended/periodic systems with greater delocalization, the block structure may broaden if canonical orbitals are used. In such cases, our framework can be combined with standard localization strategies (localized molecular orbitals for molecules or maximally localized Wannier functions~\cite{Marzari:2012aa} in periodic settings) to strengthen locality and thus further improve sparsity and numerical efficiency. For example, we have previously employed maximally localized Wannier functions and the resulting block-sparse Wannier Hamiltonian to study charge transport in MXene nanodevices~\cite{Zhou:2018aa}.

\section{Summary}
\label{sec:summary}

In summary, we have formulated and analyzed a mixed BS-THC representation of Cholesky tensors for AFQMC, leading to a cubic scaling AFQMC method across 1D/2D/3D molecular ensembles.
The central findings, a) linear NNZ growth in $L^\gamma$, b) sublinear but unsaturated average rank, and c) pronounced rank heterogeneity, are robust across dimensions and consistent with locality and inhomogeneity in molecular ensembles.
A cost model yields a \emph{size-independent} rank threshold that splits the Cholesky set into BS and THC subsets.
The resulting mixed BS-THC scheme capitalizes on both: BS bounds the per-$\gamma$ cost by $\mathcal{O}(N^2)$, and THC (with a \emph{size-independent} rank cutoff) captures inexpensive low rank contributions without incurring super-cubic growth.
The resulting AFQMC exchange-energy evaluation scales as $\mathcal{O}(N^3)$ with reduced memory, and the method preserves accuracy at practical thresholds. This enables predictive AFQMC studies of large molecular ensembles and cavity-modified chemistry in the collective regime.

Beyond AFQMC, we believe the same mixed BS-THC representation can be applied to any method whose bottlenecks are ERI-amplitude contractions. In coupled cluster, e.g., Coupled Cluster Singles and Doubles (CCSD) and Equation-of-Motion CCSD (EOM-CCSD), and perturbative corrections, the costly terms built from ERI $(ij|ab)$ and related intermediates can be rewritten as sums over Cholesky tensors; routing genuinely low rank vectors to THC while keeping the remainder block sparse reduces prefactors and, in localized bases, can lower the effective scaling on realistic systems without sacrificing accuracy.
More broadly, our framework is complementary to other existing low-rank ERI approaches~\cite{Parrish:2012aa, Dong:2018aa, Weigend:2002aa}. Low-rank structure can be introduced either at the level of the four-index ERI (e.g., alternative ERI factorizations) or via nested decompositions of the Cholesky tensors $L^\gamma$ produced by CD; the proposed mixed BS–THC strategy belongs to the latter class, and can be combined with upstream ERI decompositions and/or downstream compression/acceleration of selected $L^\gamma$ subsets to further reduce prefactors while retaining accuracy.
Besides, our framework is complementary to other exchange-acceleration strategies, and could be further integrated with alternative efficient schemes, e.g., mixed deterministic–stochastic representations for exchange \cite{Bradbury:2023aa}—potentially yielding additional reductions in wall time for large-scale calculations.
Hence, we believe our proposed mixed BS-THC scheme provides a scalable, accuracy-preserving, and broadly transferable route to cubic-scaling exchange-energy evaluation, enabling predictive simulations of large molecular ensembles and cavity-modified chemistry with AFQMC and beyond.

\begin{acknowledgments}
We acknowledge support from the US DOE, Office of Science, Basic Energy Sciences, Chemical Sciences, Geosciences, and Biosciences Division under Triad National Security, LLC (``Triad'') contract Grant 89233218CNA000001 (FWP: LANLECF7). This research used computational resources provided by the Institutional Computing (IC) Program and the Darwin testbed at Los Alamos National Laboratory (LANL), funded by the Computational Systems and Software Environments subprogram of LANL's Advanced Simulation and Computing program. LANL is operated by Triad National Security, LLC, for the National Nuclear Security Administration of the US Department of Energy (Contract No. 89233218CNA000001).
\end{acknowledgments}

\bibliography{qmconly, polariton}

\end{document}